\documentclass[prl,twocolumn,superscriptaddress]{revtex4}

\usepackage{comment}
\usepackage{amsmath,amssymb}
\usepackage{verbatim}
\usepackage{graphicx}
\usepackage{mathrsfs,yfonts}
\usepackage{latexsym}
\usepackage{amsthm}

\usepackage{amssymb}
\usepackage{hyperref}
\usepackage{color}
\usepackage{footmisc}
\usepackage{epstopdf}

\newcommand{\gsim}{\lower.7ex\hbox{$\;\stackrel{\textstyle\rangle}{\sim}\;$}}
\newcommand{\lsim}{\lower.7ex\hbox{$\;\stackrel{\textstyle\langle}{\sim}\;$}}

\definecolor{summersky}{cmyk}{0.71,0.33,0,0.14}
\definecolor{flamingo}{cmyk}{0,0.51,0.71,0.14}

\begin{document}

\title{Dynamical Tidal Locking Theory: \\
A new source of the Spin of Dark Matter Halos}

\author{E. Ebrahimian}
\email{e.ebrahimian@physics.sharif.edu}
\affiliation{Department of Physics, Sharif University of Technology, Tehran, Iran}

\author{A.A. Abolhasani}
\email{abolhasani@ipm.ir}
\affiliation{Department of Physics, Sharif University of Technology, Tehran, Iran}

\begin{abstract}
We revisit the question of what mechanism is responsible for the spins of halos of dark matter. The answer to this question is of high importance for modeling galaxy intrinsic alignment, which can potentially contaminate current and future lensing data. In particular, we show that when the dark matter halos pass nearly by each other in dense environments--  namely halo assemblies-- they swing and spin each other via exerting mutual tidal torques. We show that this has a significant contribution to the spin of dark matter halos comparable to that of calculated by the so-called tidal torque theory (TTT). We use the results of state-of-the-art simulation of Illutris to check the prediction of this theory against the simulation data.
\end{abstract}

\maketitle

\date{\today}

\preprint{...}

\section{Introduction}


Recently, there has been an ever-increasing interest in the angular momentum of the structures.
 Galaxy lensing and cosmic shear data that used to infer many cosmological parameters are sensitive to the alignment of the structures even in the dark sector \cite{Schaefer:2008xd,Troxel:2014dba,Vlah:2019byq,Fortuna:2020vsz}. The dark energy equation of state, dark matter density, neutrino mass, and spatial curvature are among the valuable data that can be achieved by lensing observations\cite{Joudaki:2017zdt}. Therefore a complete understanding of the origin of the angular momentum of the dark matter halo is highly crucial for removing a possible source of error is the cosmic shear due to intrinsic alignment \footnote{ Intrinsic alignments of galaxies can play an essential role in ellipticity correlations on small scales and need to be modeled appropriately. Commonly, one attributes intrinsic alignments of elliptical galaxies to direct interaction of the galaxy with tidal fields and traces alignments of spiral galaxies back to the orientation of their angular momentum, which in turn is related to the tidal torquing process.}.  Moreover, there is even some hope to find signatures of the primordial gravitational waves within these data \cite{Biagetti:2020lpx}. Henceforth, we should first answer the question of how structures gain angular momentum, and more importantly, how it is related to their environment.

Besides these observational impacts on the field, galaxy angular momentum is a crucial part of the structure formation theory \cite{Gamow(1952),vonWeizsacker(1951),Hoyle(1951)}. It was first pointed out by Hoyle that during gravitational collapse, the matter gain their angular momentum from an ambient gravitational field\cite{Hoyle(1951)}. The idea came to a culminating point by the seminal paper of Peebles \cite{Peebles(1969)}. He rigorously estimated the amount of angular momentum that overdense regions acquire during the collapse in the linear regime using perturbation theory. Later on, Doroshekich \cite{Doroshkevich(1970)} correctly pointed out that Peebles's result is based on the unnecessary assumption that over-dense regions are initially spherically symmetric. By relaxing this condition, he predicted that angular momentum grows linearly with time which was subsequently confirmed by computer simulations\cite{White:1984uf,Barnes:1987hu,Sugerman:1999au,Porciani:2001db,Lopez:2018lnz,Zjupa:2016xpk}. Since then, this description, known as tidal torque theory (TTT), has become the common lore, and it has been widely believed that TTT is barely enough to estimate the angular momentum of a halo. 

TTT states that angular momentum arises from tidal torques acting on an infalling matter of proto-halos or proto-galaxies during the early stage of their evolution. Arguably, the same discussion can be used to explain the spin of halos of dark matter. Based on the Zel'dovich approximation \cite{Zeldovich:1969sb}\cite{White:1984uf} one can write down the particle physical position of matter, $\mathbf{r}$, in terms of the Lagrangian comoving coordinate, $\mathbf{q}$, and Newtonian potential, $\Phi$, 
\begin{equation}
    \mathbf{r}=a(t) [\mathbf{q}-D_+(t)\mathbf{\nabla}_q \Phi]
\end{equation}
where $D_+(t)$ is the growth function and represents time evolution in comoving space. Assume a Lagrangian volume $V_L$ where contains all matters that will eventually end up in a single halo. One can express the angular momentum of this volume relative to the center of mass, $\int d^3\mathbf{r}\rho \mathbf{r}\times\mathbf{v}$ in Zel'dovich approximation as
\begin{equation}
    L(t)=-a^5\Dot{D}_+ \int_{V_L}d^3\mathbf{q}(\mathbf{q}-\mathbf{q}_{cm})\times\mathbf{\nabla}\Phi.
\end{equation}
By expanding the potential, $\Phi$, around the center of mass $\mathbf{q}_{cm}$ up to second-order we get
\begin{equation}
    \Phi(\mathbf{q})=\Phi(\mathbf{q}_{cm})+\Delta q^i\frac{\partial\Phi}{\partial q^i}\Bigg|_{q_{cm}}+\Delta q^i\Delta q^j \frac{\partial^2\Phi}{\partial q_i \partial q_j}\Bigg|_{q_{cm}}
\end{equation}
where the second term is related to the center of mass acceleration of the halo, and the third term is the tidal field. The angular momentum vector up to second order in $\Delta q$ will be
\begin{equation}
    L_i(t)=-a^2\Dot{D}_+\,\epsilon_{ijk} \frac{\partial^2\Phi}{\partial q^l \partial q^k}\Bigg|_{q_{cm}} \int_{V_L} d^3\mathbf{q}\,\rho(t)a^3 \Delta q^j \Delta q^l.
\end{equation}
In the above expression, the integral is the inertia tensor of the infalling matter in comoving Lagrangian coordinate, $\mathbf{I}$, and the potential derivative is the tidal tensor, $\mathbf{T}$, that are both time-independent in linear approximation and the time dependence only comes from overall $a^2\dot{D}_{+}$ coefficient. Note that the angular momentum vector depends on ambient matter distribution only via tidal tensor.
Assuming that $V_L$ is a sphere inertial tensor would be proportional to identity, so angular momentum vanishes to the first order \cite{Peebles(1969)} \footnote{Peebels also assumed an Eulerian spherical volume which causes the angular momentum from matter convection, see \cite{White:1984uf}}. However, according to the standard picture of structure formation, dark matter halos are associated with peaks in the initial Gaussian random field of smoothed matter density that are generally non-spherical \footnote{Generally the probability distribution of ellipticity the peaks is a function of peak height-- through $\nu = \delta_c/\sigma(R)$-- such that the higher peaks are more symmetric compared to the lower ones.} .(see also \cite{Bardeen:1985tr,Press:1973iz}).

In a matter-dominated universe one can show that $a^2\dot{b}=t$ so
$
L_i(t)=-t \epsilon_{ijk} I_{jl}T_{lk}
$
. As a result, according to TTT in the matter-dominated era, most of the proto-halos gained angular momentum by first-order tidal torque, $L(t)\propto t$ till the expansion of matter halted at the turn-around and linear theory approximation breaks down. Soon after turn-around time, the halo collapses and becomes virialized, and then its angular momentum is conserved. The final angular momentum of the halo approximately is therefore
\begin{equation}
    L_i=-t_m \epsilon_{ijk} I_{jl}T_{lk}
\end{equation}
where $t_m$ is the turn-around time. This analytic expression is very useful to investigating TTT predictions by means of simulations.

Since the amplitude of angular momentum can vary over few orders of magnitude, to quantify the angular momentum of structures, it is customary to define dimensionless spin parameter as
\begin{equation}
\lambda:=\frac{L\sqrt{|E|}}{G M^{5/2}}
\end{equation} 
in which $E$ is the energy of the halo. In simulations, there are usually some difficulties in estimating the total energy of a halo. To get around this problem, people use some alternative definitions, for example, see \cite{2014JKAS...47...77A}. According to TTT, the spin parameter, $\lambda$, is almost independent of halo mass and redshift \cite{Steinmetz:1994nb}. However spin parameter does depend on peak height  $\nu = \delta_c/\sigma(R)$ where $\sigma(R)$ is the variance of the initial matter density smoothed on the scale $R$, such that the spin parameter decreases with increasing $\nu$. It is because those high peaks are more symmetric comparing to low ones, so less net torque is exerted on them \cite{Bardeen:1985tr,Steinmetz:1994nb}. These predictions are partially supported by the simulations,  while there have been significant discrepancies between TTT predictions and simulations \cite{Barnes:1987hu,Sugerman:1999au,Porciani:2001db,Lopez:2018lnz,Zjupa:2016xpk}. Simulations confirm the prediction of TTT that the angular momentum of proto-halos grows linearly with time, and the spin parameter is almost independent of halo mass and redshift (depending on spin parameter definition). Moreover,  the result of the simulations, as one may expect, reveals that the spin parameter has a well known log-normal distribution with $\Bar{\lambda}\approx0.4-0.7$.  Nevertheless, the amplitude and direction of the spin generally have deviations from TTT prediction.  That should not be a surprise since the TTT works only in the linear limit, whereas halos are highly non-linear objects.

Numerous studies have investigated the correlations between the spin of halos and their environment 
\cite{Wang:2018rlf,Wang:2017okt,Zhang:2014rju,2012MNRAS.427.3320C}. This has crucial impacts on the interpretation of LSS surveys like galaxy lensing and cosmic shear, \cite{Schaefer:2008xd,Troxel:2014dba,Vlah:2019byq,Fortuna:2020vsz,Natarajan:2001xc,Giahi-Saravani:2013cse,Schmitz:2018rfw,Codis:2014awa,Crittenden:2000wi}. These correlations can be explained if we know the exact mechanism of angular momentum acquisition.

There are two options to go beyond the standard TTT : (i) extending TTT  to the non-linear era \cite{Porciani:2001db,Laigle:2013tsa,2015MNRAS.449.2910C} and (ii) taking halo mergers into account \cite{Vitvitska:2001vw,Bett:2015aoa,Drakos:2019tel,2020arXiv200109208B}. However, in the case of mergers, the same tidal field which, accelerates matter in the linear era, also accelerates halos in the merger process, so both mentioned mechanisms are within gravitational instability picture \cite{2017ASSL..430..249S}.

There is still another non-linear mechanism that can significantly alter the final angular momentum of halos but has not been considered yet, and we are to elaborate it this letter. This new mechanism works in a very different way than TTT or mergers, and has nothing to do with ambient matter field; when two halos pass by each other, they can change their spins by a process  pretty much like the tidal locking process. Tidal locking has been widely studied in our Solar System, and other binary or N-body systems \cite{Murray}, the tidal torques acting on a spheroidal object changes its angular momentum until they become locked together, like The Moon to The Earth.  \footnote{There are many more examples in Solar System and even beyond. The Earth is going to be locked to the Moon, and the Mercury's rotation is greatly affected by the Sun, however, due to its somewhat elliptic orbit, full tidal locking has not occurred for the Mercury.}  That is a known effect for dark matter halos\cite{Pereira:2007sw}, however, to the best of our knowledge,  the effect of mutual tidal locking of halos on their angular momenta have not been considered yet while it has a high impact on relating spin of halos to their environment. We show that tidal locking has a considerable effect on the angular momentum of halos. The rest of this paper organized as follows: In the next section, we will study the acquisition of angular momentum during close encounters of halo, dubbed Tidal Locking Theory (Hereafter TLT). Afterward, we show that the predictions of this theory follow the result of state-of-the-art simulations --like Illustris-- to an unprecedented accuracy level.

\section{Tidal Locking Theory}
The essence of tidal locking theory is that a similar mechanism that locked the Moon to the Earth is partly responsible for the spin of dark matter halos. However, there are differences between the Moon-Earth locking scenario and dark matter halo tidal locking. In most binary systems like the Earth-Moon system, which are tidally locked to each other, two objects are compact spheres that orbit in circular paths around each other for a long time and tidal forces of objects causes them to reshape. As a result of the difference between orbital and rotation periods, induced tidal bulges will be misaligned with the tidal field, so mutual gravitational pull exerts a torque. This bilateral torque changes the angular momentum of the masses, and eventually, in long enough time, they will become tidally locked. 

\begin{figure}
    \includegraphics[scale=0.45]{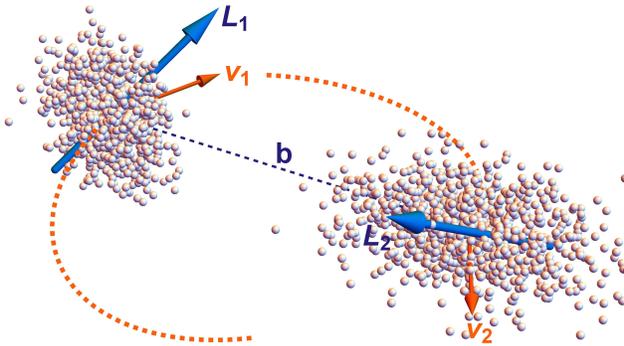}
    \caption{Two halos that passing by each other can transfer their orbital angular momentum to intrinsic angular momentum.}
    \label{fig:my_label}
\end{figure}

However, over-dense regions of dark matter are statistically asymmetric \cite{Bardeen:1985tr}, so their mutual gravitational pull exert torques once they are closely passing each other, see FIG \ref{fig:my_label}. While in this scenario, there is not enough time for a complete tidal locking-- like what happened to our Moon-- it can give rise to a substantial change in the angular momentum of halos. Contrary to the planets that are too compact and highly spherical, halos of dark matter are in general, extended diffuse non-symmetric, so even a short encounter is enough to change the angular momentum to a great extent.

Without going much to the details, let us estimate the angular momentum acquisition in a single close encounter. Assume two halo with masses $M_1$ and $M_2$ passing by each other with relative speed $v$ and impact parameter $b$. Maximum value of the tidal tensor of $M_1$ (actually largest eigenvalue) is like $GM_2/b^3$ and inertial tensor of $M_1$ is about $\frac{2}{5}M_1 R_1^2$ and encounter time is about $b/v$ so the angular momentum gained by $M_1$  can be estimated as
\begin{equation}\label{kikspin}
\Delta L_1\approx \frac{2}{5} \frac{GM_1 M_2 R_1^2}{b^2 v}
\end{equation}
By performing a detailed analysis we find very similar expression, however we ignore the details and assume that
\begin{equation}\label{kikspin2}
\Delta L_1= f\frac{GM_1 M_2 R_1^2}{b^2 v_0}
\end{equation}
in which $f$ is a factor of order unity which can be determined by fitting to the simulation data.

Before moving on to the results, let us estimate the number of close encounter events per Hubble time to emphasize their importance. The number of all encounter events for a single halo in a Hubble time is
\begin{equation}
N\sim \pi b^2 n v H^{-1}.
\end{equation}
Let us define a close encounter as an event where two halos get closer than $n^{-1/3}$ which $n$ is their environment number density, then
\begin{equation}
N\sim \pi n^{1/3} v H^{-1}
\end{equation}
where $v$ is the typical speed of halos. To get an estimate by assuming $v=400\,\mathrm{km}/\mathrm{s}$ and $n\sim 1\mathrm{Mpc}^{-3}$ one would get $N\sim 10$.  It means that the events of close encounters are frequent enough to affect the angular momentum of halos.

An important remark is in order here.  We know that halos of dark matter do not remain intact until today, and they underwent mergers and smoothly accrete mass. Therefore,  we need to first separate the angular momentum gain due to those mass changes from that is due to the tidal locking effect. While dark matter particles accreting onto a halo, they gain angular momentum from the ambient gravitational field and eventually convey it to the host halo. Let us consider a halo with a mass $M$, which accrete mass of $\Delta M$. The exact amount of the angular momentum gained by every $\Delta M$ is unknown, but statistically, the total angular momentum gained will not change the spin parameter, so we call this angular momentum $\Delta L_{\mathrm{TTT}}$ and
\begin{equation}\label{DLTTT}
        \langle\Delta L_{\mathrm{TTT}}\rangle=\bar{\lambda}\langle\Delta ( \sqrt{2 GM^3 R})\rangle
\end{equation}
which $\bar{\lambda}$ is mean spin parameter. We claim that the residue angular momentum from subtracting $\Delta L_{\mathrm{TTT}}$ from total angular momentum change is due to the tidal locking effect. We will discuss this in more detail in the following.

\begin{table}[t]
    \centering
    \begin{tabular}{ll}
    \hline
        $L_{\rm box}$ (comoving) & 106.5 Mpc \\
        $m_{DM}$ & $7.5\times 10^6 M_{\odot}$\\
        $N_{DM}$ & $1820^3 $\\
        $N_{\rm snap}$ & 136 \\
        $\Omega_{m}$ & 	0.2726\\
        $\Omega_{\Lambda}$ & 0.7274\\
       $ \Omega_{bar}$ & 0\\
        h & 0.704\\
        $N_{\rm subfind}(z=0)$& 4872374\\
        \hline
    \end{tabular}
    \caption{Illustris-1-Dark simulation}
    \label{tab:my_label}
\end{table}

\section{Data and analysis}
\label{Data_And_Analysis:sec}

When it comes to studying the complex non-linear  systems of particles with gravitational interaction, N-body simulations are always unrivaled tools. Significant advances have been made in cosmological simulations over the past decades in both size and resolution. The Illustris simulations are a series of state-of-the-art large cosmological simulations that, for their huge halo counts, are suitable for studying statistics of galaxies and dark matter halos. The results of the simulation are well organized and easy to work with \footnote{Thanks to The web-based interface (API) provided by the Illustris team, there is no need to download the whole data at each snapshot.}, which is highly crucial for dealing with such amount of data. In this simulation, halos and subhalos \footnote{In the Illustris simulation, halos are the largest gravitational structures made of the coalition of smaller sub-structures--called sub-halos. In simple words, halos are a group of subhalos} are identified by the FoF algorithm and merger trees constructed by both SubLink and LhaloTree algorithms. Subhalo mass, spin, speed, radius, etc. are among the data available in this simulation.  Also, the most massive progenitor of each subhalo at a previous snapshot (based on merger tree) and data of the host group of the subhalo can be read rather quickly from the data. We use Illustris-1-Dark (dark matter only) simulation to study predictions of TLT. Noteworthy that thanks to the well-organized data structure of data of Illustris project, one can repeat this analysis on its other simulations. Some additional technical details of the Illustris-1-Dark simulation are shown in Table \ref{tab:my_label}.

To calculate the angular momentum induced by the tidal locking mechanism, we need to know subhalo radius, $R$, mass, $M$, speed, $v$, mean mass of neighboring subhalos, $\bar{M}$, and mean number density of subhalos in the environment,$n$. Subhalo mass and speed are directly available in the simulation data. We use the radius containing half of the total mass as the radius of subhalo, $R$, this radius is also directly available in the simulation. To find the mean mass of neighboring subhalos, we divide the host halo mass by the number of subhalos in that host halo: $\Bar{M}=M_{\mathrm{Host-group}}/N_{\mathrm{subfind}}$. The mean number density of the environment that subhalo lies within can be defined as the number of subhalos in the host halo divided by the volume of the sphere around the host halo whose mean density is 200 times the critical density of that snapshot, $n=3N_{\mathrm{subfind}}/4\pi r_{200}^3$.

\begin{figure}
   \includegraphics[scale=0.35]{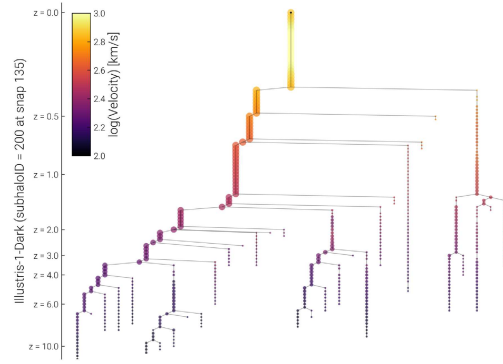}
    \caption{Typical merger tree from Illustris-1-dark simulation, the size of the circles are scaled with the mass}
    \label{fig:Mergertree}
\end{figure}

For the current analysis, we use 3000 most massive subhalos of the most massive group at the $z=0$, the mass ranges from $2\times10^9 M_{\odot}$ to $3\times 10^{14} M_{\odot}$. Massive subhalos are considered to reduce the low-particle noise effect \cite{2010ApJ...711.1198T}. After that, we track back these subhalos by their most massive progenitor. One can see a typical merger tree in the FIG.\ref{fig:Mergertree}, actually this is the subhalo number 200 in the snapshot number 135 (final snapshot) created by SubLink. Tracking back through the most massive progenitor at each snapshot can be read from the leftmost branch, the size of the circle shows the mass.

Here, we focus on the change amplitude of the angular momentum of every halo in two successive snapshots. At each snapshot, we calculate the net \emph{angular impulse} on a halo with subtracting the angular momentum vector of the halo from the angular momentum vector of its most massive progenitors in the previous snapshot so at each step we have
\begin{equation}
\Delta \vec{L}=\Delta\vec{L}_{\mathrm{TTT}}+\Delta\vec{L}_{TLT}
\end{equation}
where $\Delta\vec{L}_{\mathrm{TTT}}$ is the change of angular momentum  predicted by TTT, Eq.\eqref{DLTTT}, and $\Delta \vec{L}_{\mathrm{TLT}}$ is the angular momentum change due to the tidal locking process. Sourced by independent mechanisms, directions of the $ \Delta\vec{L}_{\mathrm{TTT}}$  and $\Delta\vec{L}_{\mathrm{TLT}}$ are uncorrelated so average of the squared angular impulse for each snapshot is therefore
\begin{equation}\label{eq2}
\langle\Delta L^2\rangle=\langle\Delta L_{\mathrm{TTT}}^2\rangle+\langle\Delta L_{TLT}^2\rangle.
\end{equation}
Now we  find $\langle\Delta L_{TLT}^2\rangle$ by subtracting the TTT part from total change and then compare it with our analytical estimate given in Eq.\eqref{kikspin2}. We find $\langle\Delta L^2_{\mathrm{TTT}}\rangle$ by considering the fact that the spin parameter in TTT is almost independent of the mass and redshift, and the mean magnitude of the spin parameter is $\langle\lambda_{\mathrm{TTT}}\rangle\approx 0.45$. This assumption is supported by both simulations and analytical calculations \cite{Steinmetz:1994nb}, so by choosing $\bar{\lambda}=0.45$ in Eq.\eqref{DLTTT} we can find $\Delta L_{\mathrm{TTT}}$. Remind that this assumption is valid only in statistical sense and for the subhalos that are virialized. We introduce $\langle\Delta L_{\mathrm{TLT}}^2\rangle_{\mathrm{sim.}}$ as 
\begin{equation}
    \langle\Delta L_{\mathrm{TLT}}^2\rangle_{\mathrm{sim.}}=\langle\Delta L^2\rangle-\langle\Delta L_{\mathrm{TTT}}^2\rangle
\end{equation}
which can be directly read from the simulation. 

On the other hand, $\langle\Delta L^2_{TLT}\rangle$ can be easily calculated by the tidal locking theory and the data provided in the simulation. The angular momentum change due to the tidal locking effect is like a random walk which after $N$ steps one would have: $\Delta X^2\approx l^2 N$ where $\Delta X^2$ is the distance from the origin and $l$ is the length of each step, so for a step duration like $\Delta t$ we have
\begin{equation}
\label{LTLT:eq}
\langle\Delta L_{TLT}^2\rangle_{\mathrm{th.}}=f^2 \bigg\langle\int \left(\frac{GM_1\bar{M} R_1^2}{b^2 v_0}\right)^2 2\pi b \, \mathrm{d}b\  v_0 n \Delta t\bigg\rangle
\end{equation}
for each subhalo in the simulation we assume that $v_0$ is the subhalo velocity, $n$ is the mean number density of the host group, $\bar{M}$ is the mean mass of subhalos in the host group, $R_1$ is the radius of subhalo and $\Delta t$ is the time step between two snapshots. Integral over impact parameter $b$ is taken from the radius of subhalo, $R_1$, to mean distance of the subhalos in the host group, $n^{-1/3}$ -- above which there are so many other subhalos that their effects average to zero. The result of integral on $b$ is like $n^{2/3}-R_1^{-2}$, so it is relatively robust to slight changes in integration limits. We calculate the median of these parameters among 3000 subhalos at each snapshot and report them in the result section. Using the median of these parameters has the advantage of neglecting violent events among the data. Note that since the data varies over a few orders of magnitudes, the mean of the trajectories does not produce a soft curve to compare the results.

Now everything is at hand  to read $\langle\Delta L_{TLT}\rangle_{\mathrm{th}}$ and $\langle\Delta L_{TLT}\rangle_{\mathrm{sim}}$ from the simulation and compare them with each other. Besides we find $f$ by fitting $\langle\Delta L_{TLT}\rangle_{\mathrm{th}}$ to $\langle\Delta L_{TLT}\rangle_{\mathrm{sim}}$ form the data.
\begin{figure}[h]
\includegraphics[scale=0.6]{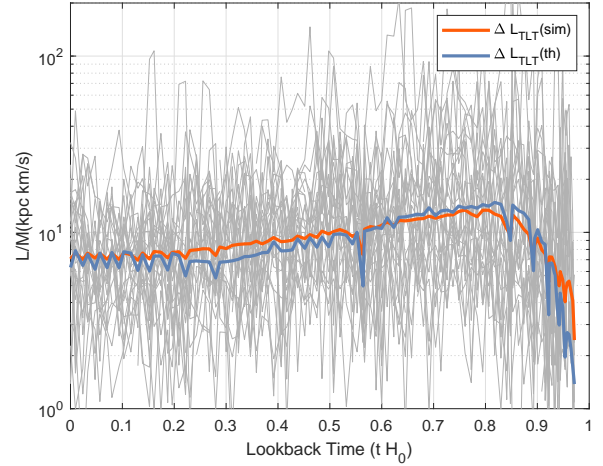}
\caption{Angular momentum vs Time, ripples are due to the non-uniform time steps used in the simulation}\label{fig1}
\end{figure}

\section{results and outlook}

We investigate the total angular impulse (net change of angular momentum) in successive snapshots and compare it with the prediction of the Tidal Locking Theory. Our main result appears in Fig.\ref{fig1}. The average angular momentum change of the 3000 subhalos due to the tidal locking is plotted versus the lookback time.  One hundred consecutive snapshots of the Illustris simulation used to get this result. The red curve is the $\langle\Delta L_{TLT}\rangle_{\mathrm{sim}}$ per mass unit and the blue curve show the prediction of tidal locking theory  $\langle\Delta L_{TLT}\rangle_{\mathrm{th}}$ given by Eq.\eqref{LTLT:eq}. The blue curve,$\langle\Delta L_{TLT}\rangle_{\mathrm{th}}$, obtained by $f\approx 0.26$ which is compatible with our estimations. The exact value of $f$ can be found, in principle, once we know the statistics of the halo shapes. Note that our rough estimation at \ref{kikspin} suggest that $f \simeq 0.4$. Grey lines at the background angular momentum change of a bunch of subhalos $|\Delta L_{\mathrm{sim}}|$-- as a sample-- plotted versus time.
The vertical axis of the plot is in  logarithmic scales. The typical value of angular impulse in a successive snapshot is $\sim 10^0-10^2\,\mathrm{kpc\,km\,s^{-1}}$. At the same time, the median-- the red  line-- vary only of the order of $\sim 10 \mathrm{kpc\,km\,s^{-1}}$. Moreover, in terms of the dimensionless spin parameter, we find that $\Delta\lambda_{\mathrm{TLT}}\approx0.02-0.03$, while $\lambda_{TTT}\approx 0.05$, that means the tidal locking effect is strong enough to change the direction of the halo spins significantly from that of predicted by TTT. Noteworthy that the small ripples on both red and blue curves are artifacts that are caused by the fact that for some technical reasons, time steps between two successive snapshots are not uniform.

 The prediction of tidal locking theory proposed in this paper is in a remarkable agreement with the simulation. We found a new mechanism of angular momentum acquisition of halos, which has a contribution comparable with that of TTT. Moreover, this finding has crucial impacts on the interpretation of observation of forthcoming LSS surveys. As mentioned earlier, weak-lensing observations are sensitive to the angular momentum of the structures, \cite{Schaefer:2008xd,Troxel:2014dba,Vlah:2019byq,Fortuna:2020vsz,Natarajan:2001xc,Giahi-Saravani:2013cse,Schmitz:2018rfw,Codis:2014awa,Crittenden:2000wi}. Lensing surveys are becoming more and more precise. They contain valuable data; however, to extract this data, we need a model for galaxy alignment correlations, which is dependent on the angular momentum. In our model, the angular momentum of halos does depend on their environment, so we provide a new explanation for observed spin-LSS correlations. In future studies, we will move on to investigate spin-LSS correlations in more detail, using the environment effects on velocities, and halo’s  initial alignment can lead us to explain the spin-LSS correlations along with TTT and mergers.

\section*{Acknowledgement}
We would like to thank Shant Baghram, Mohammad Ansari Fard, Mehrdad Mirbabayi, Sadra Jazayeri, Bjoern Malte Schaefer,  Jorge Noreña, and Sohrab Rahvar for their fruitful comments and discussions. We also want to thank Farnik Nikakhtar, who kindly helped us with the DarkSky simulation. We are also greateful to the the TNG team for providing a web-based interface to work with the data and also Dylan Nelson for his helpful guides during working with the data.

\end{document}